\documentclass[]{aa}

\pdfoutput=1					

\usepackage{graphicx}
\usepackage{amssymb}
\usepackage[latin1]{inputenc} 		
\usepackage{natbib}
\bibpunct{(}{)}{;}{a}{}{,} 			
\usepackage{txfonts} 			

\usepackage{xcolor} 			
\usepackage{hyperref}			
\hypersetup{
    pdfborder=0 0 0,     			
    pdftitle={Hard X-ray identification of Eta Carinae and steadiness close to periastron},
    pdfauthor={Leyder J.-C., Walter R. and Rauw G.},
    pdfsubject={INTEGRAL observations of the hard X-ray emission from Eta Carinae close to and far from periastron},
    pdfkeywords={X-rays: binaries, X-rays: individuals: Eta Carinae, X-rays: individuals: 1E 1048.1-5937, X-rays: individuals: IGR J10447-6027},
}

\begin{document}

\title{Hard X-ray identification of $\eta$~Carinae\\ and steadiness close to periastron}

\author{
J.-C.~Leyder \inst{\ref{IAGL},\ref{ISDC}} \fnmsep\thanks{FNRS Research Fellow} \and
R.~Walter \inst{\ref{ISDC},\ref{ObsGE}} \and
G.~Rauw \inst{\ref{IAGL}}  \fnmsep\thanks{FNRS Research Associate}
}

\offprints{J.-C. Leyder}

\institute{
Institut d'Astrophysique et de Géophysique, Université de Liège, Allée du 6-Août 17, Bâtiment B5c, B--4000 Liège, Belgium \label{IAGL} \\ \email{leyder@astro.ulg.ac.be} \and
ISDC --- Data Centre for Astrophysics, Université de Genève, Chemin d'Écogia 16, CH--1290 Versoix, Switzerland \label{ISDC} \and
Observatoire de Genève, Université de Genève, Chemin des Maillettes 51, CH--1290 Sauverny, Switzerland \label{ObsGE}
}

\date{Received 24 February 2010 / Accepted 17 August 2010}

\abstract
{
The colliding-wind binary $\eta$~Carinae exhibits soft X-ray thermal emission that varies strongly around the periastron passage. It has been found to have non-thermal emission, thanks to its detection in hard X-rays using \textit{INTEGRAL} and \textit{Suzaku}, and also in $\gamma$-rays with \textit{AGILE} and \textit{Fermi}.
}
{
This paper attempts to definitively identify $\eta$~Carinae as the source of the hard X-ray emission, to examine how changes in the 2--10 keV~band influence changes in the hard X-ray band, and to understand more clearly the mechanisms producing the non-thermal emission using new \textit{INTEGRAL} observations obtained close to periastron passage.
}
{
To strengthen the identification of $\eta$~Carinae with the hard X-ray source, a long \textit{Chandra} observation encompassing the \textit{INTEGRAL}/ISGRI error circle was analysed, and all other soft X-ray sources (including the outer shell of $\eta$~Carinae itself) were discarded as likely counter-parts.
To expand the knowledge of the physical processes governing the X-ray lightcurve, new hard X-ray images of $\eta$~Carinae were studied close to periastron, and compared to previous observations far from periastron.
}
{
The \textit{INTEGRAL} component, when represented by a power law (with a photon index $\Gamma$ of 1.8), would produce more emission in the \textit{Chandra} band than observed from any point source in the ISGRI error circle apart from $\eta$~Carinae, as long as the hydrogen column density to the ISGRI source is lower than $N_{\mathrm{H}} \lesssim 10^{24}$~cm$^{-2}$.
Sources with such a high absorption are very rare, thus the hard X-ray emission is very likely to be associated with $\eta$~Carinae. The eventual contribution of the outer shell to the non-thermal component also remains fairly limited.
Close to periastron passage, a 3-$\sigma$ detection is achieved for the hard X-ray emission of $\eta$~Carinae, with a flux similar to the average value far from periastron.
}
{
Assuming a single absorption component for both the thermal and non-thermal sources, this 3-$\sigma$ detection can be explained with a hydrogen column density that does not exceed $N_{\mathrm{H}} \lesssim 6 \times 10^{23}$~cm$^{-2}$ without resorting to an intrinsic increase in the hard X-ray emission.
The energy injected in hard X-rays (averaged over a month timescale) appears to be rather constant at least as close as a few stellar radii, well within the acceleration region of the wind.
}



\maketitle

\section{Introduction}
\label{sec:Introduction}
Located in the Carina nebula at a distance of 2.3~kpc \citep{Smith-06}, \object{$\eta$ Carinae} is not only one of the most massive objects of our Galaxy (80--120~$M_{\sun}$; \citealt{Davidson+97, Hillier+01}), but also one of the most peculiar and impressive. During its ``Great Eruption'', it briefly became the second brightest object in the sky in 1843, before fading to a V magnitude $m_{\mathrm{V}}$ of 8. A second notable eruption took place in 1890. In the middle of the XXth century, it started to slowly and irregularly brighten again, up to its current value of $m_{\mathrm{V}} \simeq 5$ \citep[see e.g.][]{Davidson+99,Fernandez-Lajus+10}.

The extended bipolar nebula that can be seen around $\eta$~Car, the so-called \textit{homunculus}, is due to the huge quantities of matter (10--20~$M_{\sun}$; \citealt{Smith+03}) ejected during the 1840s, while the \textit{little homunculus} is due to the ejection of approximately $0.1$~$M_{\sun}$ during the second burst in the 1890s \citep{Ishibashi+03, Smith-05}. The circumstellar gas that now enshrouds the object makes observations in the optical and ultra-violet energy ranges very difficult. $\eta$~Car continues to eject matter through energetic stellar winds, with an estimated mass-loss rate of $10^{-4}$--$10^{-3}$~$M_{\sun}$\,yr$^{-1}$ \citep{Andriesse+78, Hillier+01, Pittard+02-EtaCar, vanBoekel+03}.

\subsection{$\eta$~Carinae as a binary system}
\label{subsec:binary}
Although many questions remain open, there is now strong evidence that $\eta$~Car is a binary system. For example, a period of $\simeq 5.5$~yr, which has been stable over decades, was inferred from observations in the radio \citep{Duncan+95}, millimeter (mm; \citealt{Abraham+05}), optical \citep{Damineli-96, Damineli+00}, near-infrared (near-IR; \citealt{Whitelock+94, Whitelock+04, Damineli-96}), and X-ray \citep{Corcoran-05} domains. \citet{Damineli+08-periodicity} found an orbital period of $\sim2022.7 \pm 1.3$~days (5.53~yr) by combining radial velocity variations from multiple lines in optical spectra along with broad-band IR, optical, and X-ray observations.

The primary star is most likely a luminous blue variable (LBV; \citealt{Davidson+97}), a short phase during the evolution of massive stars that occurs after leaving the main sequence and leads to the object becoming a Wolf-Rayet star. The secondary star is probably a late-type nitrogen-rich O or WR type \citep{Iping+05, Verner+05}. \citet{Mehner+10} derived a zone in the Hertzsprung-Russel diagram where the secondary is most likely to lie (see their Fig.~12). The orbit has a semi-major axis of $16.64$~AU \citep{Hillier+01} and is very eccentric ($e \simeq 0.9$; \citealt{Nielsen+07}). These values lead to a periastron distance $r_{\mathrm{periastron}}$ of 1.66~AU, which should be compared to the (poorly known) primary star's radius $R_{\star, 1}$, believed to be $\simeq 0.7$~AU \citep{Corcoran+07}, but which might be as high as 1~AU \citep{Damineli-96}.

The orientation of the binary system remains unclear. A number of studies have concluded that the secondary moves behind at periastron \citep{Damineli-96, Pittard+98, Corcoran+01, Corcoran-05, Akashi+06, Hamaguchi+07, Nielsen+07, Kashi+08-orientation, Okazaki+08, Parkin+09}, but others seem to favour the opposite situation \citep{Falceta-Goncalves+05, Kashi+07}, while quadrature has also been suggested \citep{Ishibashi-01, Smith+04}.

\subsection{X-ray observations}
\label{subsec:XRayObservations}
Observations of $\eta$~Car in the X-ray domain are relatively unaffected by absorption compared to most other wavelengths, thus allowing an investigation deep into the system \citep[see][]{Corcoran+07}.

First shown by the \textit{Einstein} satellite \citep{Chlebowski+84}, and later confirmed by \textit{Chandra} observations \citep{Seward+01}, the X-ray emission of $\eta$~Car has two distinct components:\\
-- a softer, spatially extended (up to about 20\arcsec), and inhomogeneous thermal ($kT_{\mathrm{SX}} \simeq 0.5$~keV) X-ray component (named $\eta$\,SX), which dominates the spectrum mostly below 1.5~keV, and is probably associated with the interaction between the stellar wind and the interstellar matter;\\
-- a harder, point-like thermal ($kT_{\mathrm{HX}} \simeq 4.7$~keV) X-ray component ($\eta$\,HX), centered on the stellar system, that dominates the spectrum in the 2--10~keV range and is probably linked to the wind collision between the two massive stars that form the binary system.

The X-ray spectrum is coherent with a colliding-wind binary (CWB) interpretation \citep{Usov-92, Stevens+92, Pittard+97, Pittard+98, Ishibashi+99, Pittard+02-EtaCar, Corcoran-05, Pittard+07, Parkin+09}. In this framework, the slow and dense stellar wind from the LBV ($v_{\infty, 1} \simeq 500$~km/s, $\dot{M}_{1} \simeq 2.5 \times 10^{-4}$~$M_{\sun}$/yr; \citealt{Pittard+02-EtaCar}) collides with the higher-velocity, lower-density wind from the hotter and luminous companion star ($v_{\infty, 2} \simeq 3000$~km/s, $\dot{M}_{2} \simeq 1.0 \times 10^{-5}$~$M_{\sun}$/yr; \citealt{Pittard+02-EtaCar}). The X-ray emitting region where the hydrodynamical shock takes place is called the wind-wind collision region (WCR).

The X-ray lightcurve of $\eta$~Carinae is of particular interest : it exhibits impressive variations, with the orbital periodicity, and undergoes an extremely deep and relatively long minimum close to periastron (see Sect.~\ref{sec:Xray-lightcurve} for details). Variations at many other wavelengths are also coincident with the periastron passage; for example, optical excitation lines undergo a rapid decrease in intensity \citep{Damineli+08-multispectral}. Although the X-ray emission represents only a tiny fraction of the bolometric luminosity ($L_{\mathrm{X}}/L_{\mathrm{bol}} \simeq 10^{-7}$), understanding the origin of the hard X-ray emission is crucial, as it is compact and not affected by circumstellar absorption. Moreover, the X-ray lightcurve exhibits short-term flares, with a quasi-period of 84~d \citep{Ishibashi+99,Corcoran-05}. \citet{Moffat+09} discuss different interpretations of these flares, favouring an explanation based on the presence of clumps in the wind of the LBV.

A hard X-ray tail (\textit{i.e.}, at a flux higher than an extrapolation of the thermal X-ray component $\eta$\,HX) was once detected in the $\eta$~Car region by the PDS instrument onboard \textit{BeppoSAX} \citep{Viotti+04}. Since then, it was unambiguously confirmed by \textit{INTEGRAL} \citep{Leyder+08}, and reobserved by \textit{Suzaku} \citep{Sekiguchi+09}. This hard X-ray detection is important, as it proves that non-thermal particle acceleration operates in the wind collision, and implies that gamma-ray emission is likely in the MeV and GeV energy ranges. The emission mechanism is believed to be inverse Compton (IC) scattering of UV photons by electrons accelerated in the hydrodynamical shock between the stellar winds \citep{Leyder+08}.

\subsection{$\gamma$-ray observations}
\label{subsec:GammaRayObservations}
In the $\gamma$-ray domain, the Carina region has been observed since July 2007, initially by the \textit{AGILE} satellite \citep{Tavani+08}. \citet{Tavani+09} report the detection of \object{1AGL J1043-5931} above 100~MeV, based on observations taken with the Gamma-Ray Imaging Detector instrument (GRID; 30~MeV--30~GeV; \citealt{Prest+03}). The flux of this source, averaged over all the observations, is $(3.7 \pm 0.5) \times 10^{-7}$~photon\,cm$^{-2}$\,s$^{-1}$ (above 100~MeV).
\citet{Tavani+09} claim that this source, located at $(l,b) = (287.6, -0.7)$ (with an error of 0.4\degr), is probably associated with $\eta$~Car (located 0.07\degr\ away). They also report that a 2-day $\gamma$-ray flare was observed in October 2008.

In the same high-energy range, the Carina region has also been monitored since August 2008 by the Large Area Telescope (LAT; \citealt{Atwood+09}) instrument onboard the \textit{Fermi} satellite. According to the \textit{Fermi} LAT first source catalog \citep{Abdo+10}, the source \object{1FGL J1045.2-5942} is located at a position of $(l,b) = (287.6252, -0.6388)$ (with a 95\%-confidence semi-major axis of 1.4\arcmin). This position is 1.7\arcmin\ away from $\eta$~Car, and the sources cannot be formally associated with each other. The flux is $(2.6 \pm 0.2) \times 10^{-10}$~erg\,cm$^{-2}$\,s$^{-1}$ (in the 100~MeV--100~GeV energy range).
In both this catalog and \citet{Tavani+09}, it is indicated that there might be an association between the \textit{AGILE} and the \textit{Fermi} sources (although they are separated by 4\arcmin).

In this paper, new hard X-ray observations performed with \textit{INTEGRAL} close to the periastron passage of $\eta$~Carinae are presented. Section~\ref{sec:Xray-lightcurve} describes the behaviour of the X-ray lightcurve of $\eta$~Car. Section~\ref{sec:Observations} deals with the data reduction and analysis of the new \textit{INTEGRAL} observations; while Sect.~\ref{sec:Results} describes the newly obtained images and spectra with respect to the previous observations in the hard X-ray domain. Section~\ref{sec:Counterpart} is focused on definitively identifying the counter-part of the \textit{INTEGRAL} source as $\eta$~Car. Section~\ref{sec:Discussion} discusses these new observations, and conclusions are finally drawn in Sect.~\ref{sec:Conclusions}.

\section{The X-ray lightcurve}
\label{sec:Xray-lightcurve}

\subsection{RXTE observations}
\label{subsec:RXTEObservations}
\begin{figure}[htbp]
\centering
\includegraphics[width=1.0\columnwidth]{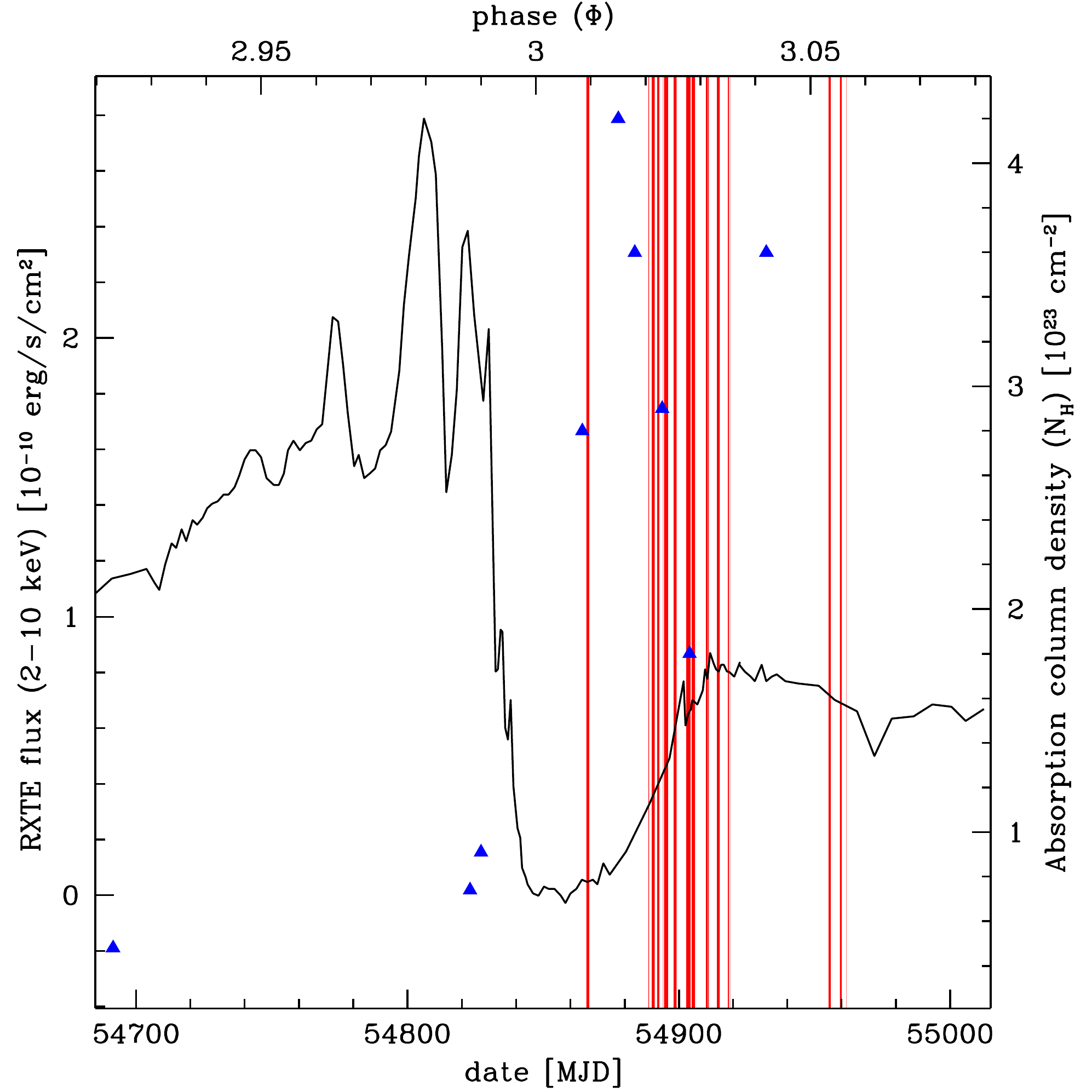}
\caption{\textit{RXTE} lightcurve of $\eta$~Carinae (in the 2--10~keV energy range), from August~2008 to June~2009, corrected for instrumental background (from Corcoran et al.\ 2010, \textit{in prep.}). The bottom axis gives the time in MJD, while the top axis gives the phase of $\eta$~Car (based on the ephemeris from \citealt{Corcoran-05}). The start of the minimum is close to the periastron passage from 2009 ($\Phi = 3$). Red lines indicate the new \textit{INTEGRAL} observations, obtained just after the X-ray minimum. Blue triangles are the $N_{H}$ measured during the previous minimum (with their axis on the right side), as reported by \citet{Hamaguchi+07}.}
\label{fig:RXTE-lightcurve}
\end{figure}

Observations obtained by the \textit{Rossi X-ray Timing Explorer} (\textit{RXTE}; \citealt{Swank-06}) since 1996 cover over two orbital cycles of $\eta$~Carinae, including the three minima from 1998, 2003.5, and 2009. The X-ray lightcurve around the last X-ray minimum is shown in Fig.~\ref{fig:RXTE-lightcurve} (see \citealt{Corcoran-05} and \citealt{Moffat+09} for details about \textit{RXTE} data analysis), with superimposed red lines indicating the new \textit{INTEGRAL} observations around periastron. The blue triangles (for which the axis is given on the right side) are the absorption column density values obtained from \textit{Chandra} and \textit{XMM-Newton} observations around the 2003 minimum, as reported by \citet{Hamaguchi+07} in their Table~4.

The general behaviour of the X-ray lightcurve is closely described by the CWB model \citep{Pittard+98, Ishibashi+99, Pittard+02-EtaCar, Corcoran-05}. Although the X-ray minimum (\textit{i.e.} phase $\Phi = 0$) probably occurs around the time of the periastron passage, both events might be separated by up to a few months \citep{Damineli+00, Corcoran+01}. The scenario is as follows: when the two stars are far from each other (\textit{i.e.} $\Phi \gtrsim 0.1$), the X-ray flux is roughly constant. After apastron ($\Phi > 0.5$), the secondary approaches the primary, the wind interactions become more important, and the X-ray flux increases steadily over approximately 2~years. Then, close to periastron, the X-ray flux drops sharply by a factor of more than $\simeq 100$ (as is shown by an \textit{XMM-Newton} observation taken during the minimum; \citealt{Hamaguchi+07}) for a duration of 80--90~days. The key physical processes explaining this sudden decrease remain uncertain. After remaining at this minimum level for approximately 3~months, the X-ray flux returns to its roughly constant value as the stars recede from each other, before starting a new cycle. The X-ray lightcurve follows the same periodicity of $2024\pm2$~days as observed at other wavelengths \citep{Ishibashi+99, Corcoran-05, Hamaguchi+07}.

\subsection{Interpretation of the X-ray minimum}
\label{subsec:Interpretation}
To explain the minimum in the X-ray emission close to periastron, different scenarios have been proposed (not necessarily mutually exclusive). There are two broad categories of proposed models: some are based on a change in the \emph{intrinsic} emission (e.g. a modification of the WCR itself), while others postulate a variation in the \emph{observed} emission only (either by occultation, or by any process leading to an increase in the column density). These two types are developed below.

The first category is based on the fading (or complete disappearance) of the WCR. In this model, as the separation between the stars decreases near periastron, they become so close that the WCR enters into the acceleration region of the secondary's stellar wind. The momentum balance becomes unstable, the primary star's wind overwhelms the other one, and the WCR collapses onto the surface of the secondary star.

The second explanation of the decrease in the observed emission is based on the eclipse model, in which the WCR becomes totally occulted around periastron. This can be caused by either the primary star itself, or its optically-thick wind; both cases require the secondary to move behind the primary. However, the minimum is too long to be explained only by an eclipse due solely to the star itself, and would not be constant for such a long period because the stellar motion close to periastron is very rapid \citep{Hamaguchi+07}.

Another explanation of the observed X-ray variation is a temporary increase in the mass-loss rate. In this hypothesis, an increase in the mass-loss rate leads to more material being ejected into the interstellar medium (ISM), and therefore a shielding of the UV photons, thus encouraging the formation of dust and an increase in absorption.

The X-ray minimum can therefore be explained by either a decrease in the plasma emission measure (EM), an increase in the column density, or a combination of those two options. The column density $N_{\mathrm{H}} \simeq 5 \times 10^{24}$~cm$^{-2}$ necessary to reduce the X-ray flux by two orders of magnitude from the maximum to the minimum is indeed much higher than the value observed ($N_{\mathrm{H}} \simeq 5 \times 10^{23}$~cm$^{-2}$; \citealt{Hamaguchi+07}), hence the need for an additional explanation, distinct from the column density variation.

Hard X-ray observations at the X-ray minimum, such as those provided by \textit{INTEGRAL}, which are mostly unaffected by absorption $N_{\mathrm{H}} \lesssim10^{24}$~cm$^{-2}$, can help distinguishing between these different models.

\section{\textit{INTEGRAL} observations and data analysis}
\label{sec:Observations}
The ESA \textit{INTEGRAL} $\gamma$-ray satellite carries 3 coaligned instruments dedicated to the observation of the high-energy sky, from 3~keV up to 10~MeV, along with an optical monitoring camera \citep{WInkler+03}. The \textit{INTEGRAL} soft gamma-ray imager (ISGRI; \citealt{Lebrun+03}), the most sensitive imaging detector between 15 and 200~keV currently flying, offered the first apparent detection of $\eta$~Car at hard X-rays \citep{Leyder+08}.

The new \textit{INTEGRAL} observations are indicated by the red lines superimposed on the \textit{RXTE} lightcurve in Fig.~\ref{fig:RXTE-lightcurve}. They were scheduled for a period covering 80~days after the expected 2009 minimum, as the previous minima from 1998 and 2003.5 lasted 80--90~days. Unfortunately, the 2009 minimum happened to be half as long as the previous two X-ray minima, explaining why only a fraction of the \textit{INTEGRAL} observations correspond to the minimum.

Two images were produced, to compare the behaviour of $\eta$~Car close to and away from periastron. The first image contains all available data located within 15\degr\ of $\eta$~Car, and which covers the minima; this includes data from the 2003.5 X-ray minimum, and from the 2009 X-ray minimum (see Table~\ref{tab:INTEGRAL-data}). The second image was compiled using all available public data located within 15\degr\ of $\eta$~Car and not included in the first image; this second image is thus comparable to the one from \citet{Leyder+08}, although the effective exposure time is significantly longer. A total of 362 (respectively 2027) pointings were added to produce the first (respectively second) image, for 
a deadtime-corrected good exposure of 1.1~Ms (respectively 5.5~Ms). ISGRI pointing sky images were produced with a pre-release of OSA\footnote{The Offline Scientific Analysis (OSA) software is available from the ISDC website~: \url{http://www.isdc.unige.ch}} version~9, using standard parameters. The image-cleaning step used an input catalogue of all \textit{INTEGRAL} sources with fixed source positions, and was applied independently of the source strength (thus allowing for negative source models) to avoid introducing any biases during the process.

\begin{table}[htdp]
\caption{\textit{INTEGRAL} data available for $\eta$~Car close to the periastron passage; only data coincident with the X-ray minimum were selected (\textit{i.e.} phase $0 < \Phi <0.04$).}
\label{tab:INTEGRAL-data}
\centering
\begin{tabular}{ccccc} \hline \hline
Minimum & Revolutions & Time [MJD] & Phase $\Phi$ of $\eta$~Car \\ \hline 
2003.5 & 88--91 & 52\,824--52\,833 & 2.000--2.005 \\
2009 & 770--788 & 54\,865--54\,920 & 3.009--3.036\\ \hline

\end{tabular}
\end{table}

Broad-band (17--80~keV) mosaic images of the field were produced, along with narrower bands in 12 energy ranges.
The final mosaic images were compiled using equatorial coordinates with a tangential projection, using an oversampling factor of two with respect to the individual sky images. The photometric integrity and accurate astrometry were ensured by calculating the intersection between input and output pixels, and weighting the count rates with the overlapping area.

\section{\textit{INTEGRAL} images and spectra}
\label{sec:Results}
The previous analysis of ISGRI data outside periastron \citep{Leyder+08} identified three sources in the field of $\eta$~Car: $\eta$~Car itself, \object{IGR J10447-6027}, and the anomalous X-ray pulsar (AXP) \object{1E 1048.1-5937}.

\begin{figure}[htbp]
\centering
\includegraphics[width=0.49\columnwidth]{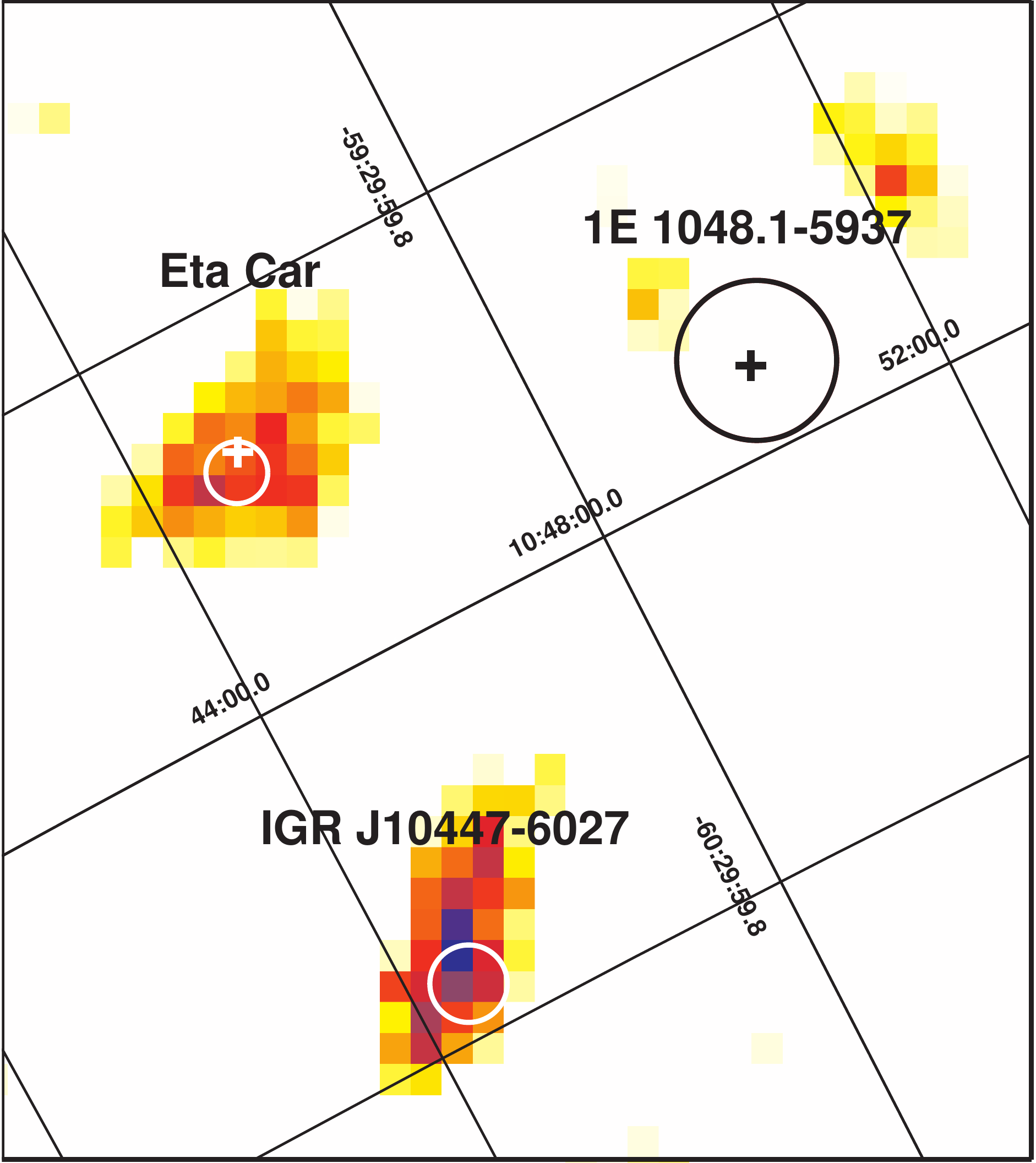}
\includegraphics[width=0.49\columnwidth]{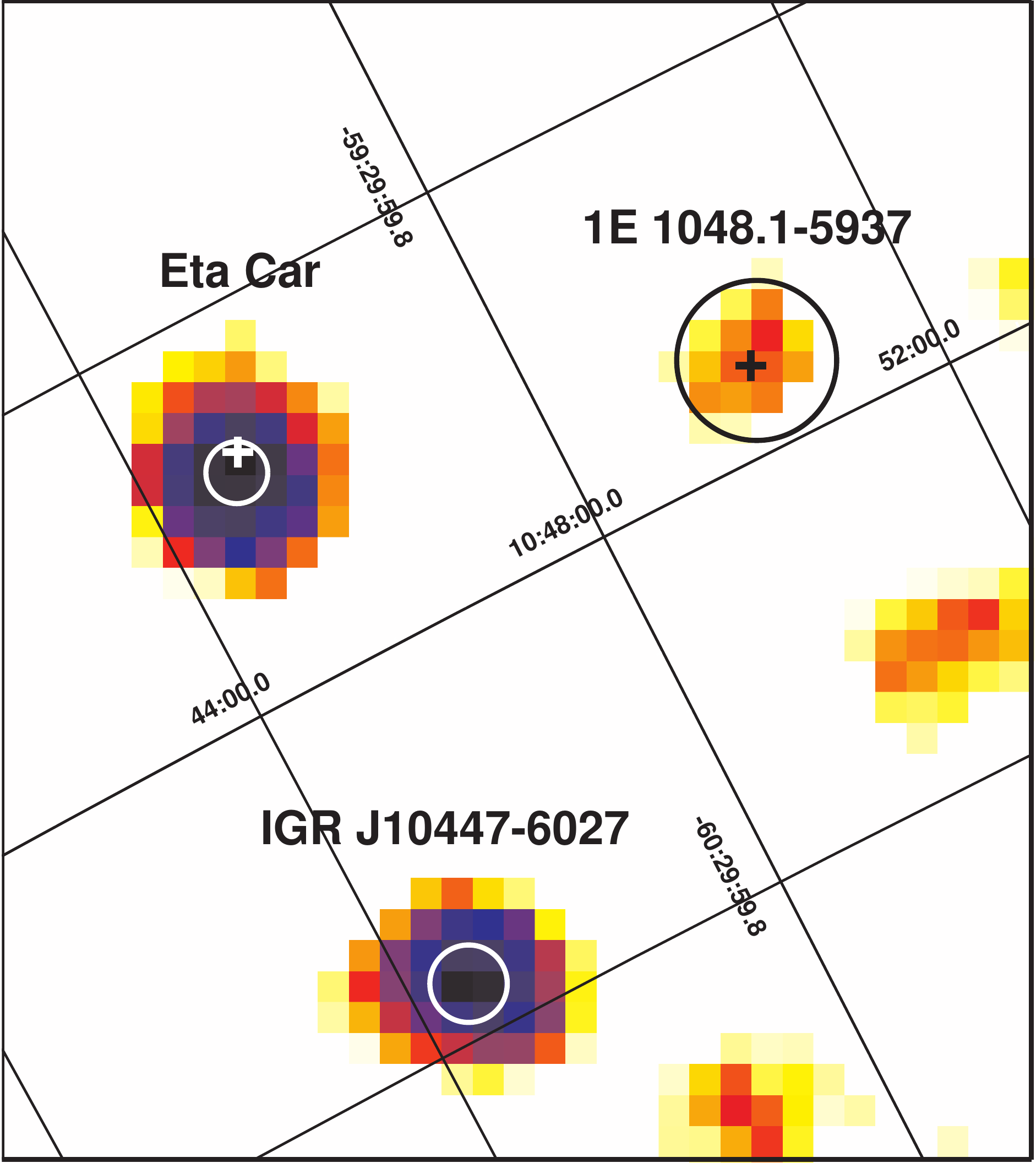}
\caption{ISGRI significance image of $\eta$~Car \emph{close to periastron} (on the left) and \emph{outside periastron} (on the right). Both images are in equatorial coordinates, cover the same energy range (17--80~keV), and use the same colourmap (significance goes from 2$\sigma$ to 7$\sigma$). The positions of  $\eta$~Car and 1E~1048.1-5937 are indicated by the white and black crosses. The white and black circles represent the best-fit positions listed in Table~\ref{tab:INTEGRAL-sources}.}
\label{fig:ISGRI-17-80}
\end{figure}

The new ISGRI 17--80~keV image \emph{close to periastron only} (\textit{i.e.}, the first image, shown in the left panel of Fig.~\ref{fig:ISGRI-17-80}) indicates that the hard X-ray emission of $\eta$~Car close to the X-ray minimum is around $0.16 \pm 0.05$~cnt\,s$^{-1}$. Although neither $\eta$~Car nor IGR~J10447-6027 are detected at a level of 5-$\sigma$, excesses are clearly observed at both positions at a level of 3-$\sigma$.

The new image \emph{far from periastron} (\textit{i.e.} the second image, shown in the right panel of Fig.~\ref{fig:ISGRI-17-80}) indicates that both $\eta$~Car and IGR~J10447-6027 are very well detected, providing more accurate positions and average fluxes than \citet{Leyder+08}. IGR~J10447-6027 is thus likely to be a persistent source. In contrast, the AXP 1E~1048.1-5937 is no longer clearly detected (although an insignificant excess is observed at its position), as opposed to the situation reported in \citet{Leyder+08}, and is therefore probably variable.

Table~\ref{tab:INTEGRAL-sources} summarizes the best-fit position, error circle radius, intensity (or 5-$\sigma$ upper limit), and significance for the sources detected outside periastron; it also indicates the intensity and significance at periastron.

\begin{table*}[htbp]
\caption{Sources detected around $\eta$~Car in ISGRI mosaics.}
\label{tab:INTEGRAL-sources}
\centering
\begin{tabular}{cccccc} \hline \hline
Source & Intensity [cnt\,s$^{-1}$] & Position (J2000) & Distance [\arcmin] & Error circle radius [\arcmin] & Significance\\ \hline 
\multicolumn{6}{c}{\textit{Close to periastron} (Fig.~\ref{fig:ISGRI-17-80}, left panel)} \\ \hline
$\eta$~Car 		& $ 0.16 \pm 0.05 $ 	& 																		 	&		& 	 	& 3.3	 	\\
IGR~J10447-6027 	& $ 0.20 \pm 0.05 $ 	& 															 				&		& 	 	& 4.2	 	\\ \hline
\multicolumn{6}{c}{\textit{Far from periastron} (Fig.~\ref{fig:ISGRI-17-80}, right panel)} \\ \hline
$\eta$~Car 		& $0.18 \pm 0.02$ 	& RA~= $10^\mathrm{h}44^\mathrm{m}57^\mathrm{s}$, Dec~= $-59\degr42\arcmin27\arcsec$ 	& 1.6		& 2.4 	& 8.4 	\\
1E~1048.1-5937 	& $< 0.11 $ 		& RA~= $10^\mathrm{h}50^\mathrm{m}14^\mathrm{s}$, Dec~= $-59\degr53\arcmin10\arcsec$ 	& 0.6		& 6.2 	& 2.9 	\\
IGR~J10447-6027 	& $0.14 \pm 0.02$ 	& RA~= $10^\mathrm{h}44^\mathrm{m}36^\mathrm{s}$, Dec~= $-60\degr25\arcmin50\arcsec$ 	& 1.9		& 3.0 	& 6.3 	\\ \hline
\end{tabular}
\tablefoot{The flux and significance values were extracted from the 17--80~keV energy band with the best-fit position, and with the PSF size set to 6\arcmin; the values close to periastron are for the 3-$\sigma$ detections; the upper-limit is at the 5-$\sigma$ level; the distance refers to the extent between the positions derived from the ISGRI observations and the published positions; the error circle corresponds to a 90\% probability. To extract the values from the observations close to periastron, the position of the sources were fixed to their best-fit positions derived from the observations far from periastron (since the lower significance of the former implies that they provide far less precise positions).}
\end{table*}

The new ISGRI best-fit position, now only 1.6\arcmin\ away, is significantly closer to $\eta$~Car than the previous one \citep{Leyder+08}, and the error circle is slightly smaller, decreasing to 2.4\arcmin. These new values strengthen the association of the \textit{INTEGRAL} source with $\eta$~Car.

The average source flux for $\eta$~Car observed by ISGRI (extracted from the second image by fixing the PSF to 6\arcmin\ and the source to its optical position) is $0.18 \pm 0.02$~count\,s$^{-1}$ in the 17--80~keV energy range. This corresponds to a flux of $1.2 \times 10^{-11}$~erg\,cm$^{-2}$\,s$^{-1}$ when assuming a photon index of 1.8 (which leads to the best fit of the data, as will be shown below). Based on a distance of 2.3~kpc \citep{Smith-06}, the hard X-ray luminosity is $8 \times 10^{33}$~erg\,s$^{-1}$ (17--80~keV).

The limited number of points in the ISGRI spectrum (extracted from data obtained far from periastron) only allows simple models to be fitted to the hard X-ray spectrum. A power law alone gives a photon index $\Gamma$ of $1.7 \pm 0.5$.
As there are no soft X-ray data simultaneous with the \textit{INTEGRAL} observation, an archival \textit{BeppoSAX} observation (from December 29, 1996) was used with the ISGRI data. Note however that this X-ray observation is adopted as a reference only, and might slightly overestimate the contribution of the thermal component in the ISGRI band. The model applied was a combination of an optically thin, thermally emitting plasma \citep{Mewe+85, Mewe+86, Kaastra-92, Liedahl+95} with a power law, both undergoing the same amount of interstellar photoelectric absorption \citep[using cross-sections from][]{Morrison+83} and Thomson scattering (\textit{i.e.} \texttt{wabs*cabs*(mekal+powerlaw)} in \texttt{Xspec} terms). The parameters derived are a photon index of $\Gamma=1.8 \pm 0.4$ and a temperature of $kT=4.3$~keV (giving a reduced $\chi^2$ of 3.6 for 78 degrees of freedom, down to 1.5 when assuming a 5\% systematic error).

This new value for the photon index is in closer agreement with \textit{Suzaku} observations than the previous value based on \textit{INTEGRAL} data \citep{Leyder+08}. The contribution of the optically thin thermal plasma emission (\texttt{mekal}) in the first ISGRI bin remains important (about 40\%), and is poorly known because of the lack of a simultaneous soft X-ray observation (\citealt{Hamaguchi+07} report variations by a factor 2 above 3~keV).

The quality of the fit can be improved by allowing the chemical abundance of the plasma to vary and by ignoring the softest part of the \textit{BeppoSAX} spectrum (which would require the addition of another optically thin thermal plasma model, \textit{i.e.} \texttt{mekal}, emitting at a different temperature, to be correctly modelled), leading to a reduced $\chi^2$ of about 2 \citep[for details, see][]{Viotti+04}; this however does not affect the values derived for the photon index.

\begin{figure}[htbp]
\centering
\includegraphics[width=1.0\columnwidth]{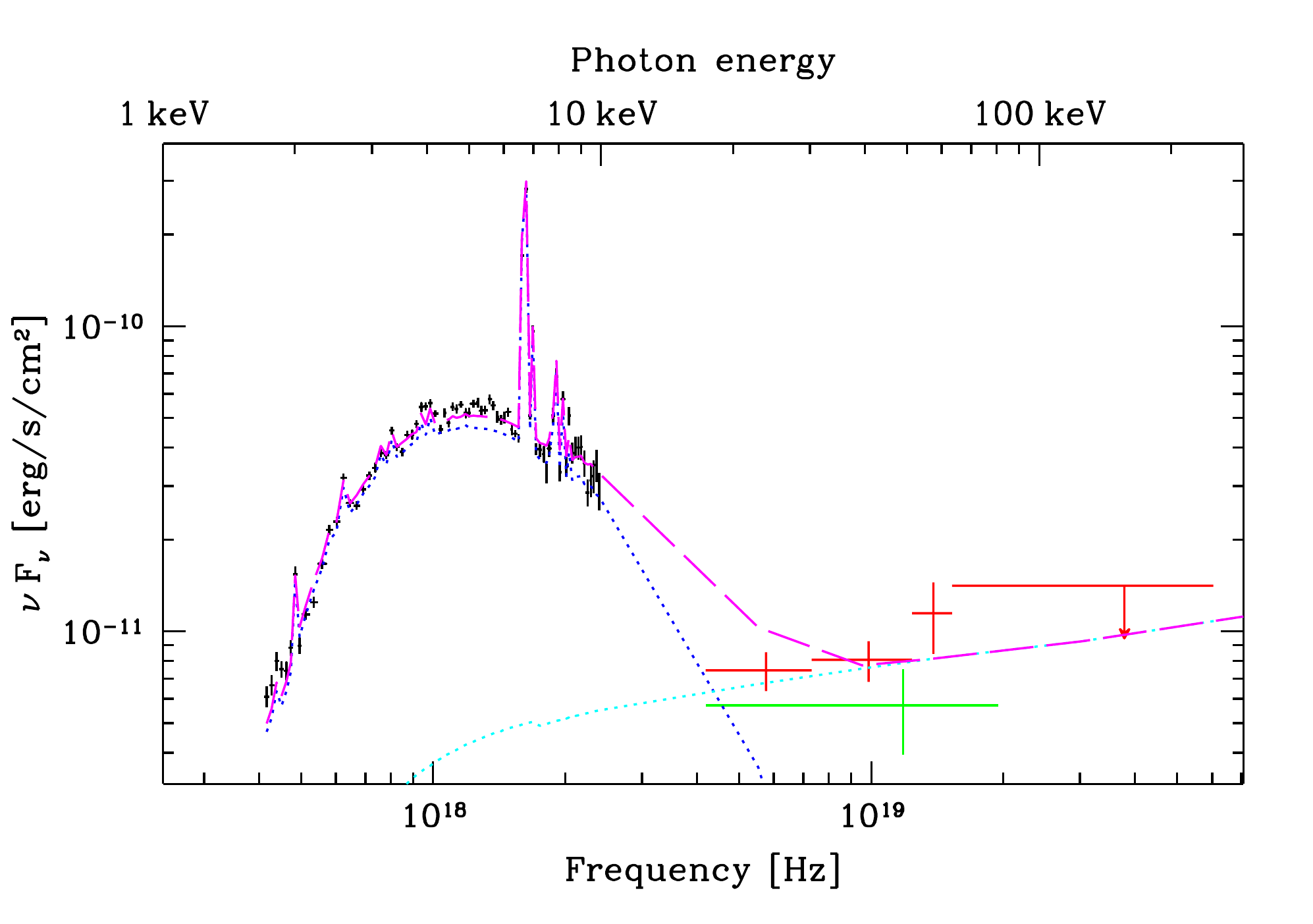}
\caption{\textit{BeppoSAX}/MECS (black) and \textit{INTEGRAL}/ISGRI (red) unfolded spectra of $\eta$~Carinae outside periastron (the last ISGRI point is a 5-$\sigma$ upper-limit); ISGRI 3-$\sigma$ detection close to periastron (green). The unfolded model is the magenta long-dashed line; with the thermal and non-thermal components respectively shown by the dotted dark blue and light blue lines.}
\label{fig:Spectrum}
\end{figure}

The unfolded spectrum outside periastron is shown in Fig.~\ref{fig:Spectrum}. The MECS observation from December 29, 1996 (at phase $\Phi = 0.83$; black points) is only meant to provide a reference spectrum of the source outside periastron. The red points are the ISGRI spectrum far from periastron. The green point is the ISGRI 3-$\sigma$ detection close to periastron. The unfolded model is shown in magenta, the thermal component being plotted in dark blue and the power law in light blue. The fluxes observed close to and outside periastron are consistent.

Finally, the data were also fitted with a model in which both thermal and non-thermal components are subject to different absorption columns (\texttt{wabs*cabs*mekal+wabs*cabs*powerlaw}). The best-fit solution was obtained with values very similar to those found when they are tied together, although the absorption on the power law is poorly constrained. However, hydrogen column densities higher than $N_{H} \gtrsim 1.5$~$\times 10^{25}$~cm$^{-2}$ can be clearly ruled out, as the hard X-rays become so absorbed that the first ISGRI bin can no longer be correctly modelled, independently of the model chosen.

\section{Counterpart of the non-thermal source}
\label{sec:Counterpart}
\citet{Leyder+08} associated the \textit{INTEGRAL} hard X-ray detection with $\eta$~Carinae, based mostly on this source being by far the most X-ray luminous in the immediate vicinity. However, all other X-ray sources within the ISGRI error box (based on a long \textit{Chandra} observation) are very unlikely to be the counterpart of the ISGRI source, as shown below.

To compare the hard X-ray spectrum to all X-ray sources located within the ISGRI error circle, the longest imaging observation of the $\eta$~Car field taken by \textit{Chandra} was chosen. It is an 88.2~ks observation taken in August 2006 (\texttt{ObsID} is 6402). This observation was taken far from periastron, which is desirable. The data were reduced using CIAO\footnote{The \textit{Chandra} Interactive Analysis of Observations (CIAO) software is available from the \textit{Chandra} X-ray Center (CXC) website~: \mbox{\url{http://cxc.harvard.edu/ciao/}}}, version 4.1.2, with standard parameters and methods. After data reduction, application of the appropriate good time intervals (GTIs), and selection of the energy range of interest (0.3--10.0~keV), the exposure time was reduced to 84.9~ks.

\begin{figure}[htbp]
\centering
\includegraphics[height=1.0\columnwidth, angle=0]{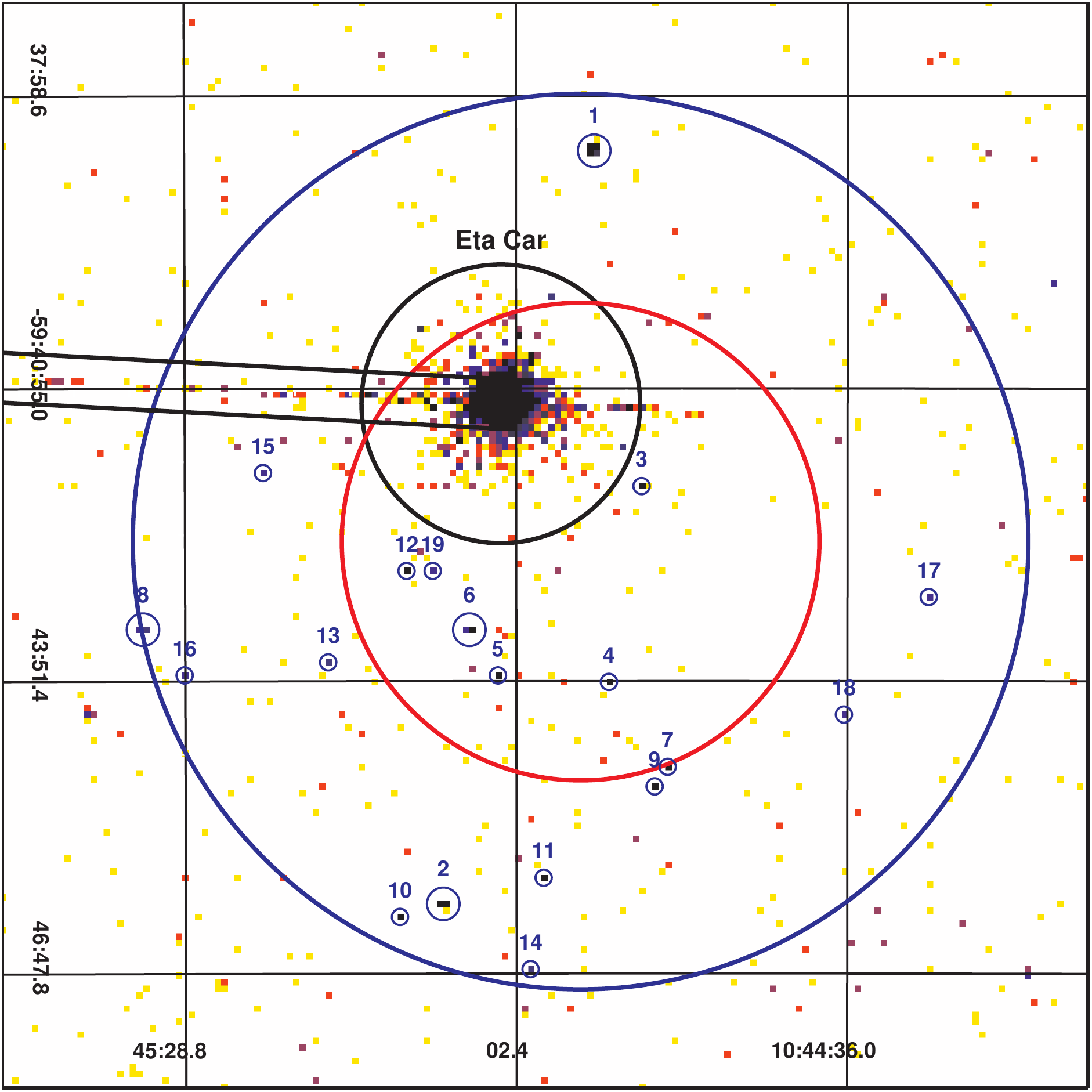}
\caption{X-ray sources detected in the \textit{Chandra} 5--10~keV image (limits range from 3 to 10~counts). The ISGRI error circle is shown in red for reference, the \textit{Chandra}-selected sources are located up to 4.5\arcmin\ from the ISGRI position (within the blue circle), and the exclusion regions (a 1.5\arcmin\ circle around $\eta$~Carinae plus a box to avoid including an artifact) are drawn in black.}
\label{fig:Chandra-sources}
\end{figure}

From a \textit{Chandra} image in the 5--10~keV energy range, all sources with more than a few counts and located inside or close to the ISGRI emission were selected (see~Fig.~\ref{fig:Chandra-sources}; the ISGRI error circle is drawn in red, while the concentric 4.5\arcmin\ selection region is shown in blue; sources are labelled in order of decreasing brightness, provided that they have more than 7~counts). Events linked to either $\eta$~Car or its nebula were not considered, \textit{i.e.} those within 1.5\arcmin\ of the source position or in the black box (which is an artifact due to the huge brightness of $\eta$~Car); both exclusion regions are indicated in black in Fig.~\ref{fig:Chandra-sources}. 
In this image, $\eta$~Car is by far the brightest object, with significantly more than ten thousand counts inside the black circle. The following sources (numbered 1, 2, and 3) contain, respectively, about 100, 50, and 40~counts.

It is known that the brightest sources in this field are associated with massive stars, while the majority of fainter sources are believed to be pre-main sequence (PMS) stars \citep[e.g.][]{Antokhin+08}. Those sources are most accurately represented by an optically thin thermal plasma emission model taking into account the interstellar photoelectric absorption (\texttt{wabs*mekal}), thus they have hardly any contribution to the hard X-ray range.

\begin{figure}[htbp]
\centering
\includegraphics[height=1.0\columnwidth, angle=0]{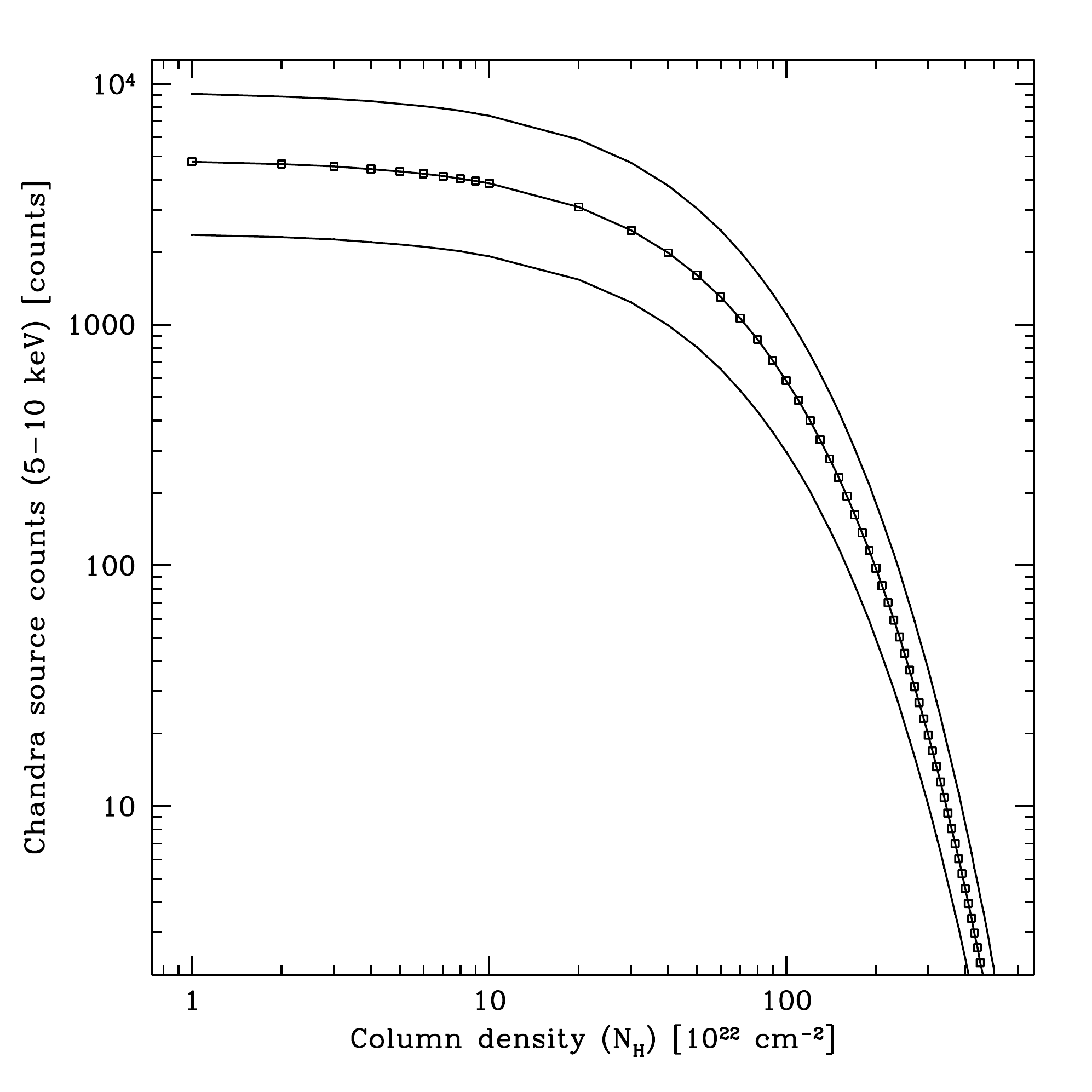}
\caption{\textit{Chandra} photon counts expected in the 5--10~keV energy range as a function of the hydrogen column density ($N_{H}$), for a photon index of $\Gamma$=1.4 (bottom curve), 1.8 (middle curve), and 2.2 (top curve). These extrapolations of the ISGRI observations are based on a power-law model that includes interstellar photoelectric absorption (\texttt{wabs*powerlaw}). }
\label{fig:ChandraCountRateVersusNH}
\end{figure}

To estimate the counts expected in the \textit{Chandra} image from the ISGRI source, its spectrum was fitted with a simple absorbed power-law model including interstellar photoelectric absorption (\texttt{wabs*powerlaw}) and extrapolated down to the \textit{Chandra} energy range (see Sect.~\ref{sec:Results}). The \textit{Chandra} source counts expected in the 5--10~keV energy range are shown in Fig.~\ref{fig:ChandraCountRateVersusNH}. The three curves correspond to three different values of the photon index ($\Gamma$=1.4, 1.8, and 2.2), thus covering the range of possible values as obtained from the ISGRI observations.

Adopting the lower limit from the range of hydrogen column densities found by \citet[][$N_{H}$ = 5--$40 \times 10^{22}$~cm$^{-2}$]{Hamaguchi+07}, because this lower value is observed during most of the orbit, the number of counts expected in the 5--10~keV energy range from the extrapolation of the \textit{INTEGRAL} spectrum is of the order of 4300. This is thus nearly a factor of three smaller than the number of counts originating from $\eta$~Car detected with \textit{Chandra} in the same energy range. In addition, as the 2--10~keV flux in this \textit{Chandra} observation is about 5 times lower than the one from the \textit{BeppoSAX} observation shown in Fig.~\ref{fig:Spectrum}, the ratio of the thermal model fitting the MECS data to the power-law extrapolation from the ISGRI data is understood.

No source other than $\eta$~Car detected with \textit{Chandra} within the ISGRI error circle has more than approximately 100~counts in the 5--10~keV energy range. According to Fig.~\ref{fig:ChandraCountRateVersusNH}, this means that if one of these other sources were the counterpart to the ISGRI emission, the hydrogen column density towards that source would have to be at least as high as $N_{H} \gtrsim 10^{24}$~cm$^{-2}$ to be compatible with both the \textit{INTEGRAL} and \textit{Chandra} data. This is a lower limit, as a source with a soft excess or a leaky absorber would require an even higher hydrogen column density to match the observations.

Such hard X-ray emission affected by a large absorption could be indicative of either a Compton-thick AGN, or a highly obscured high-mass X-ray binary (HMXB). For the latter, out of the more than 250 sources lying within the Galactic plane \citep[see the fourth ISGRI catalog from][]{Bird+10}, only one HMXB has an absorption of $N_{H} \gtrsim 10^{24}$~cm$^{-2}$. As the ISGRI spectrum extracted at the location of $\eta$~Car does not match the typical spectrum of an accreting pulsar (\textit{i.e.} a power law with a cutoff at $\sim$10~keV), this is an unlikely possibility.

It is also unlikely that a Compton-thick AGN might be the counterpart to the hard X-ray detection. \citet{Paltani+08} studied the distribution of AGNs around the quasar \object{3C 273}, for which the exposure time is comparable to that available on $\eta$~Car. From that sample, they derived a $\log N$--$\log S$ distribution for AGNs with an absorption column density $N_{H} > 10^{22}$~cm$^{-2}$ (see the blue curve in the bottom panel of their Fig.~9). They also placed an upper-limit of 24\% on the number of Compton-thick AGNs present in their sample. Given that the 20--60~keV flux for $\eta$~Car is around $8 \times 10^{-12}$~erg\,cm$^{-2}$\,s$^{-1}$ (when assuming a photon index of 1.8) and that the size of the error circle on $\eta$~Car is 2.4\arcmin, the probability of having a Compton-thick AGN coincident by chance with $\eta$~Car is strictly lower than 0.002\%.


An alternative explanation to the IC emission from the CWB was suggested by \citet{Ohm+10}. In their scenario, the acceleration of the non-thermal particles is due to the outer shell surrounding $\eta$~Car, which is detectable mostly in the 0.1--1~keV energy range \citep{Seward+01}. However, based on the long \textit{Chandra} observation, the number of counts detected in an annulus covering this shell (and excluding a point source) is of the order of 750~counts (in the 5--10~keV energy range). This value is an upper limit to the shell emission, as there remains some contribution (estimated to be at least 10\%) from the wings of the PSF of $\eta$~Car itself in that extraction region. Although there is indeed some small contribution from this shell to the X-ray spectrum, this is at least a factor of six lower than the number of counts expected from the non-thermal extrapolation from ISGRI data down to the same energy domain. Thus, the contribution of the shell to the hard X-ray emission, if any, remains fairly limited.

In summary, the most likely counterpart to the observed hard X-ray emission is $\eta$~Car, as other sources detected with \textit{Chandra} within the ISGRI error circle seem unable or very unlikely to match the flux level detected with \textit{INTEGRAL}.

\section{Discussion}
\label{sec:Discussion}
Thanks to its excellent spatial resolution, \textit{INTEGRAL} has provided the strongest evidence that the non-thermal emission comes from $\eta$~Carinae. All sources detected by \textit{Chandra} close to the ISGRI emission can be ruled out as likely counterparts, and the probability of a Compton-thick AGN being serendipitously located at the same position as Eta~Car is less than 0.002\%.

The general form of the X-ray lightcurve of $\eta$~Car close to periastron can be explained by an increase in hydrogen column density, a decline in intrinsic emission, or a combination of both. The motivation behind acquiring these new \textit{INTEGRAL} observations was to place constraints on the relative contributions of these two mechanisms.

The fraction of particles accelerated to high energies and injected into the hard X-ray tail is assumed to be constant or lower around periastron than during the remainder of the orbit. This is a reasonable hypothesis, since although the shock will be stronger closer to the star, the Compton cooling will be far more important because of the higher density of UV photons. This means that the electrons are cooled more rapidly, leading to a softer distribution and a reduced emissivity. Therefore, the expected high-energy emission is lower near periastron. Moreover, close to periastron, the stars are so close to each other with respect to the stellar radii that the winds may not have time to reach their full terminal speed, thus mitigating the effect of the stronger shock.

The ISGRI 3-$\sigma$ detection close to periastron shows a flux level consistent with the observations far from periastron. This means that the hard X-ray flux may not be lower near periastron than outside, and might even have to be intrinsically higher if the hydrogen column density effectively increased strongly during the \textit{INTEGRAL} observations.

During the spectral modelling, it was assumed that both thermal and non-thermal components are affected by the same absorption. \citet{Leyder+08} suggested that the emission mechanism explaining the hard X-ray tail is IC emission of UV photons by electrons accelerated in the hydrodynamical shock that forms between the stellar winds. If this is indeed the case, the hard X-rays are then formed in the same region as the soft X-rays, thus it is reasonable to assume that the hydrogen column density will be the same for both thermal and non-thermal components of the model. Moreover, the WCR is spatially extended and fairly non-uniform, thus making it difficult to mask the hard X-ray emission while allowing a much smaller absorption for the soft X-rays.

Simulations to estimate the 17--80~keV flux were performed for a large range of the hydrogen column density ($N_{H}$ = 1--1000~$\times 10^{22}$~cm$^{-2}$). The model used was a combination of an optically thin thermal plasma emission with a power law, both affected by the same amount of interstellar photoelectric absorption and Thompson scattering (\texttt{wabs*cabs*(mekal+powerlaw)}). The temperature was fixed to $kT=5.1$~keV and the photon index to $\Gamma=1.8$. In addition, the model was scaled to match either the average 2--10~keV flux observed by \textit{RXTE} far from periastron ($\simeq 7 \times 10^{-11}$~erg\,cm$^{-2}$\,s$^{-1}$), or the maximum flux observed just before periastron ($\simeq 3 \times 10^{-10}$~erg\,cm$^{-2}$\,s$^{-1}$). The latter choice corresponds to assuming that all the \textit{INTEGRAL} data were taken at the X-ray maximum (or during the minimum assuming it is explained solely by absorption), while the former is probably closer to the situation depicted by the data as the recovery had already started when $\eta$~Car was observed by \textit{INTEGRAL} (see Fig.~\ref{fig:RXTE-lightcurve}).

Given the hard X-ray flux observed close to periastron by ISGRI, the hydrogen column density must be smaller than $N_{H} \lesssim 3 \times 10^{24}$~cm$^{-2}$, otherwise the expected flux level cannot be matched. Without applying this additional scaling to match the maximum 2--10~keV flux observed by \textit{RXTE}, the extrapolation to the \textit{INTEGRAL} range implies a lower expected flux. Less absorption is thus needed to match the observed ISGRI flux, and the constraint on the hydrogen column density becomes even stronger, at $N_{H} \lesssim 6 \times 10^{23}$~cm$^{-2}$. This value is consistent with the highest value measured for the hydrogen column density just after periastron ($N_{H} = 4.2 \times 10^{23}$~cm$^{-2}$; \citealt{Hamaguchi+07}).

The latter result also indicates that if the hydrogen column density during the new ISGRI observations did not increase above this maximum value observed during previous periastron passages, the flux observed by ISGRI can be explained without the need for an intrinsic increase in the hard X-ray emission. The physical conditions in the shock are expected to vary along the orbit. However, the average hard X-ray flux emitted seems to remain constant at least as close to the periastron as the ISGRI observations ($\Phi$=3.02--3.035, corresponding to a separation of approximately 5--8 times the radius of the primary star).

This lack of variability is somewhat unexpected, as the contribution of the thermal component is known to vary strongly, and remains fairly important (about 40\%) in the first ISGRI bin. However, the available \textit{INTEGRAL} data do not have sufficient statistical significance to test this. One natural reason for this steadiness might be that the observations were not all taken during the X-ray minimum (see Fig.~\ref{fig:RXTE-lightcurve}). A possible alternative explanation of the lack of strong variability with the orbital phase could be that the hard X-ray emission originates, at least partly, from the shock between the strong winds of $\eta$~Car and its surroundings.

If both thermal and non-thermal components were to be affected by different absorptions, the simulations would remain valid provided that the hydrogen column density is larger for the power law than for the thermal emission. However, the upper limit derived would then apply to the hydrogen column density of the power-law component only, and that no constraints could be drawn for the soft X-ray part based on ISGRI observations.

To check whether the fraction of particles injected in the high-energy tail is constant even at the X-ray maximum or if it is lower than during the rest of the orbit, future ISGRI observations in the hard X-ray domain taken right before the X-ray minimum would be very useful. This would provide more constraints on the evolution of the intrinsic emissivity close to periastron. Longer observations performed in the X-ray minimum would also help placing a more stringent upper limit on the absorption column density. During the next X-ray minimum (in the Summer 2014), $\eta$~Car should be observable by \textit{INTEGRAL}.

On the basis of the ISGRI detection and the spectral shape, \citet{Leyder+08} predicted the source detection in the GeV energy range. The detection of $\eta$~Car happened indeed, first with \textit{AGILE} and subsequently with \textit{Fermi}, each with comparable values for the source flux. A graph of the spectral power flux of $\eta$~Car, ranging from keV to GeV, is shown in Fig.~4 of \citet{Tavani+09}. Over this wide range, the extrapolation from the \textit{INTEGRAL} energy range up to the energy domains observed by \textit{AGILE} and \textit{Fermi} with a photon index of $\simeq 1.8$ (as suggested by these new \textit{INTEGRAL} observations) suggests a flux of approximately $5 \times 10^{-11}$~erg\,cm$^{-2}$\,s$^{-1}$, a value compatible with the \textit{AGILE} measurement.

Unveiling all the mysteries related to the mechanism(s) responsible for the X-ray minimum is likely to require observations with the next generation of high-energy satellites, such as the nuclear spectroscopic telescope array (\textit{NuSTAR}) or the Japanese project \textit{ASTRO-H}.

\section{Conclusions}
\label{sec:Conclusions}

A long ($\sim$90~ks) \textit{Chandra} X-ray observation was analysed to study all sources detected in the 5--10~keV energy range around $\eta$~Carinae. Assuming that the \textit{INTEGRAL} component were represented by a power law (with a photon index $\Gamma$ of 1.8, as suggested by the new ISGRI data discussed in this paper), then this component would produce much more emission in the \textit{Chandra} band than seen for any point-like source within the ISGRI error circle except $\eta$~Car, provided that the hydrogen column density towards the ISGRI source remains modest ($N_{\mathrm{H}} \lesssim 6 \times 10^{23}$~cm$^{-2}$). The possible contribution of the outer shell to the non-thermal component also remains fairly modest (below 15\%).

If the counterpart to the \textit{INTEGRAL} emission were a source other than $\eta$~Car, then the absorption would have to be at least of the order of $N_{\mathrm{H}}\gtrsim 10^{24}$~cm$^{-2}$.
Such an object with a huge absorption may be either a Compton-thick AGN serendipitously coincident with $\eta$~Car (but the probability of this is below 0.002\%), or a highly obscured high-mass X-ray binary (but only one source with the enormous hydrogen column density required is currently known, and the hard X-ray spectral shape observed with ISGRI does not match that of an accreting pulsar).

Thus, the hard X-ray emission observed with \textit{INTEGRAL} is very likely to be associated with $\eta$~Carinae.

The 3$\sigma$-detection of $\eta$~Car in the hard X-ray domain close to periastron can be explained by an increase in the column density (reaching values of $N_{\mathrm{H}} \simeq 6 \times 10^{23}$~cm$^{-2}$, consistent with the values derived from observations during previous periastron passages by \citealt{Hamaguchi+07}), without the need for a strong modification of the intrinsic emission. The situation is different in soft X-rays, where the emission cannot be explained only by the observed variations in the $N_{H}$. The energy injected in hard X-rays averaged over a long (about a month) timescale seems to be quite constant even when the companion is within the acceleration region of the wind.

\begin{acknowledgements}
J.-C.L. thanks the FNRS and the Patrimoine ULg for the travel grants that made this work possible. J.-C.L. and G.R. acknowledge support from the \textit{XMM/INTEGRAL} PRODEX contract and from the Communauté Française de Belgique -- Action de recherche concertée -- Académie Wallonie-Europe. The authors would like to thank the referee for numerous helpful suggestions. 
\end{acknowledgements}

\bibliographystyle{aa}
\bibliography{Article-Final}


\end{document}